\documentclass[twocolumn,twoside,slac]{revtex4}
\usepackage{graphicx}
\usepackage{fancyhdr}
\usepackage{multirow}
\usepackage{amsmath}
\begin{document}

\title{The QCDOC supercomputer: hardware, software, and performance}

\author{P.A. Boyle$^{a,b}$, C. Jung$^{a,c}$, and T. Wettig$^{d,e}$ for
  the QCDOC Collaboration\footnote{The QCDOC collaboration also
    includes D. Chen, A. Gara (IBM Research), N.H. Christ, C.
    Cristian, S.D. Cohen, Z. Dong, C. Kim, L. Levkova, X. Liao, G. Liu,
    R.D.  Mawhinney, A.  Yamaguchi (Columbia University), B. Joo (U.
    of Edinburgh), S.  Ohta (RIKEN-BNL Research Center), R. Bennet, D.
    Stampf, and K. Petrov (BNL).}  }  
\affiliation{$^a$Department of
  Physics, Columbia University, New York,
  NY 10027, USA\\
  $^b$Department of Physics and Astronomy, University of Edinburgh,
  Edinburgh EH9 3JZ, Scotland\\
  $^c$Physics Department, Brookhaven National Laboratory, Upton, NY
  11973, USA\\
  $^d$Department of Physics, Yale University, New Haven, CT
  06520-8120,
  USA\\
  $^e$RIKEN-BNL Research Center, Upton, NY 11973, USA}

\begin{abstract}
  An overview is given of the QCDOC architecture, a massively parallel
  and highly scalable computer optimized for lattice QCD using
  system-on-a-chip technology.  The heart of a single node is the
  PowerPC-based QCDOC ASIC, developed in collaboration with IBM
  Research, with a peak speed of 1 GFlop/s.  The nodes communicate via
  high-speed serial links in a 6-dimensional mesh with
  nearest-neighbor connections.  We find that highly optimized
  four-dimensional QCD code obtains over 50\% efficiency in cycle
  accurate simulations of QCDOC, even for problems of fixed
  computational difficulty run on tens of thousands of nodes.  We also
  provide an overview of the QCDOC operating system, which manages and
  runs QCDOC applications on partitions of variable dimensionality.
  Finally, the SciDAC activity for QCDOC and the message-passing
  interface QMP specified as a part of the SciDAC effort are discussed
  for QCDOC.  We explain how to make optimal use of QMP routines on
  QCDOC in conjunction with existing C and C++ lattice QCD codes,
  including the publicly available MILC codes.
\end{abstract}

\maketitle

\thispagestyle{fancy}

\section{Introduction}

Lattice QCD directly performs the path integral for the QCD Lagrangian
by Monte-Carlo integration on a computer.  Space-time is discretised
on a 4-torus, and a large number of snapshots of typical vacuum
configurations is used to evaluate hadronic correlation functions
non-perturbatively. The numerical integration scheme introduces
finite-volume, discretisation, and statistical errors that can be
removed with sufficient compute power: lattice QCD is systematically
improvable.

With improved actions and extrapolation in the lattice spacing, fairly
modest lattice sizes, such as $32^3\times 64$, are believed to be
adequate for controlling the discretisation and finite volume effects
on most hadronic observables.

Algorithms for including the effects of quark loops in the selection
of typical vacuum configurations are numerically very expensive.  The
expense is believed to grow with a large power of the inverse quark
mass for current best algorithms, and has thus far not been paid in
full by lattice simulations, either choosing to ignore such effects
(quenching) or simulating with artificially large quark masses.  The
goal for lattice QCD is to simulate in a region where the dynamical
quark mass at least reliably connects to chiral perturbation theory,
if not to the physical masses, and requires at least many tens of
Teraflop years of computer power.

The notable characteristic of this problem is the need to focus ever
more compute power on reducing the quark mass for a fixed problem
size, rather than scaling up the problem size as more compute power
becomes available.  To this end QCDOC has been designed to allow
efficient distribution of a single lattice QCD simulation over a very
large multi-dimensional grid of a few tens of thousands of compute
nodes.  

At such extreme scalability the sparse matrix inversions
involved require both global summation and nearest neighbour
communication at a much shorter timescale than on smaller MPPs.
Consequently both global summation time and nearest neighbour latency
bite much harder, and both the QCDOC hardware and software have been
designed to address these issues very effectively, with an order of
magnitude improvement over traditional cluster technology on these key
operations.

\section{QCDOC hardware}

\subsection{Overview}

Continuing advances in the microelectronics industry have made it
possible to integrate almost all components that make up a computer
system on a single chip.  This is known as system-on-a-chip
technology.  Using this technology, the individual processing nodes in
a massively parallel computer can be greatly simplified.  This is the
idea behind the design of the QCDOC supercomputer: the processing
elements consist of a single application-specific integrated circuit
(ASIC) and an industry-standard DDR memory module.  Large machines can
then be build by simply adding many such processing elements.

The main ingredients of the QCDOC ASIC are
\leftmargini 5mm
\begin{itemize}\itemsep 0mm
\item 500 MHz, 32-bit PowerPC 440 processor core
\item 64-bit, 1 GFlops floating-point unit
\item 4 MBytes of on-chip memory (embedded DRAM)
\item controllers for embedded and external memory
\item nearest-neighbor serial communications unit with aggregate
  bandwidth of 12 Gbit/s in 12 independent directions
\item other components such as Ethernet controller, interrupt
  controller, etc.
\end{itemize}

In the following subsections, the QCDOC hardware is described in more
detail (see also Ref.~\cite{lat01}).  The main advantages of the QCDOC
design are
\begin{itemize}\itemsep 0mm
\item high scalability: 50\% sustained performance for typical
  applications on machines with several 10,000 nodes
\item low price-performance ratio of \$1 per sustained MFlops
\item low power consumption
\item high reliability
\end{itemize}

\subsection{The QCDOC ASIC}

\begin{figure*}[-t]
  \includegraphics[width=160mm]{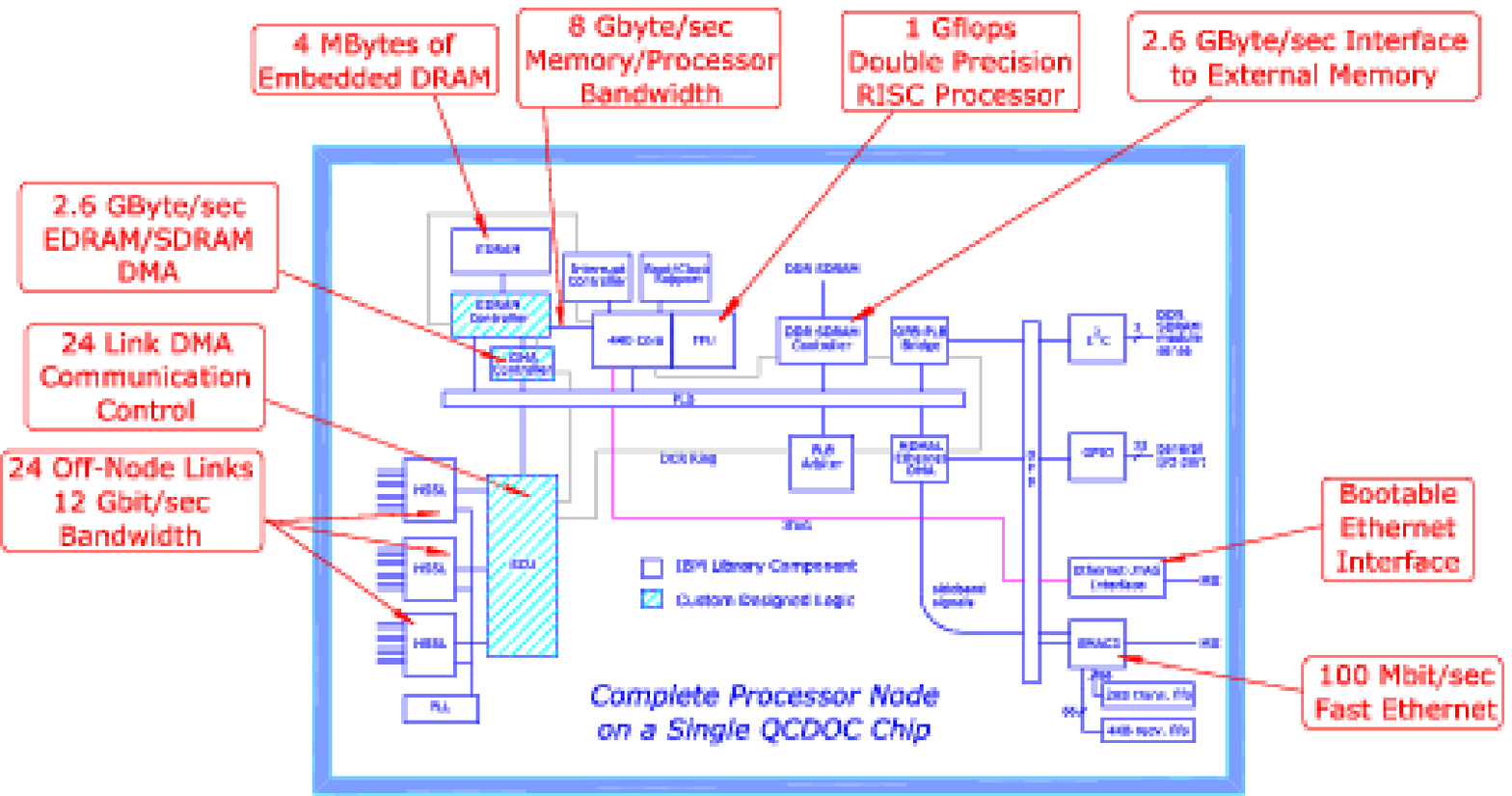}
  \caption{The QCDOC ASIC.}
  \label{fig:ASIC}
\end{figure*}

IBM is the leading vendor in the system-on-a-chip industry.  The QCDOC
ASIC was designed in close collaboration with IBM Research (Yorktown
Heights) and is shown schematically in Fig.~\ref{fig:ASIC}.  
It contains standard cores from the IBM library as well as
custom-designed components for faster memory access and high-speed
off-chip communication.  

The on-chip components are linked by three busses from IBM's
CoreConnect$^{\rm TM}$ technology \cite{PLB}:
\begin{itemize}\itemsep 0mm
\item Processor local bus (PLB).  This is a 128-bit wide, fully
  synchronous bus running at 1/3 of the CPU frequency.  It contains
  independent read and write data busses.  The PLB protocol
  (implemented by the PLB arbiter) is rather sophisticated, allowing
  for pipelining, split transactions, burst transfers (fixed and
  variable length, early termination possible), DMA transfers,
  programmable arbitration, and other features such as timeout, abort,
  etc.
\item On-chip Peripheral Bus (OPB).  This is a 32-bit wide, fully
  synchronous bus running at 1/6 of the CPU frequency.  Its basic
  purpose is to off-load slower devices from the PLB bus.
\item Device Control Register Bus (DCR).  This is a very simple bus
  which is used to read and write various control registers in the
  on-chip devices.  The 440 is the only master on this bus.
\end{itemize}

The IBM library components in the ASIC are
\begin{itemize}\itemsep 0mm
\item The PowerPC-based 440 CPU core \cite{440} with attached 64-bit
  IEEE floating point unit \cite{FPU}.  The 440 is a Book E compliant
  32-bit processor with a 32 kByte prefetching instruction cache and a
  32 kByte data cache with flexible cache control options (64-way
  associative, partitionable, lockable).  The processor includes
  memory management with a 64-entry translation-lookaside-buffer which
  supports variable page sizes from 1 kByte to 256 MByte.  It also
  features dynamic branch prediction and a 7-stage, dual issue
  pipeline.  The target frequency of the 440 is 500 MHz, i.e. the peak
  performance is 1 GFlops.\\
  The 440 also features a JTAG interface.  JTAG (Joint Test Action
  Group) is an industry-standard protocol that allows an external
  device to take complete control of the processor.  This
  functionality will be used for booting and debugging, see below.
\item 4 Mbytes of embedded DRAM (or EDRAM) which is accessed with low
  latency and high bandwidth through a custom-designed controller, the
  PEC, see below.
\item The PLB arbiter provides programmable arbitration for the up to
  eight allowed masters that can control PLB transfers.  We are using
  six masters: the 440 instruction read, data read, and data write
  interfaces (the last two are channeled through the PEC), the EDRAM
  DMA, the SCU DMA, and the MAL DMA used by the Ethernet controller.
\item The PLB-OPB bridge is used to transfer data between the two
  busses.  It is the only master on the OPB and a slave on the PLB.
\item The Universal Interrupt Controller (UIC) processes the
  interrupts that are generated on- and off-chip and provides critical
  and non-critical interrupt signals to the 440.
\item The DDR controller is a slave on the PLB, capable of
  transferring data to/from external DDR (double data rate) SDRAM at a
  peak bandwidth of 2.6 GBytes/s.  It supports an address space of 2
  GBytes and provides error detection, error correction, and refresh
  of the off-chip SDRAM.
\item The Ethernet media access controller (EMAC) provides a 100
  Mbit/s Ethernet interface (it also supports Gbit-Ethernet, but we
  are not making use of this capability).  The media-independent
  interface (MII) signals at the ASIC boundary are connected to a
  physical layer chip on the daughterboard.  The EMAC is a slave on
  the OPB and has sideband signals to the MCMAL on the PLB.
\item The DMA-capable Memory Access Layer (MCMAL) loads/unloads the
  EMAC through the sideband signals.  It is a master on the PLB.
\item The inter-integrated circuit (I$^2$C) controller is a slave on
  the OPB.  It is used to communicate with off-chip devices supporting
  the I$^2$C protocol, such as the serial presence detect EPROM on the
  DDR DIMM or voltage and temperature sensors on the motherboard.
\item The general purpose I/O (GPIO) unit is another slave on the OPB
  whose 32-bit wide data bus is taken out to the ASIC boundary.  It is
  used, e.g., to drive LEDs, to control the global interrupt tree, and
  to receive interrupts from off-chip devices.
\item The high-speed serial links (HSSL) used by the SCU each contain
  eight independent ports (four each for send/receive) through which
  bits are clocked into/out of the ASIC at 500 Mbit/s per port.  The
  bits are converted to bytes (or vice versa) in the HSSL.  The HSSL
  input clocks are phase-aligned by another IBM macro, the
  phase-locked loop (PLL).
\end{itemize}

In addition to the components provided by IBM, the QCDOC ASIC contains
custom-designed components.  Most importantly, these components
provide essential support for the high-speed communications required
in a massively parallel machine.

  \begin{figure*}[t]
    \includegraphics[angle=270,width=150mm]{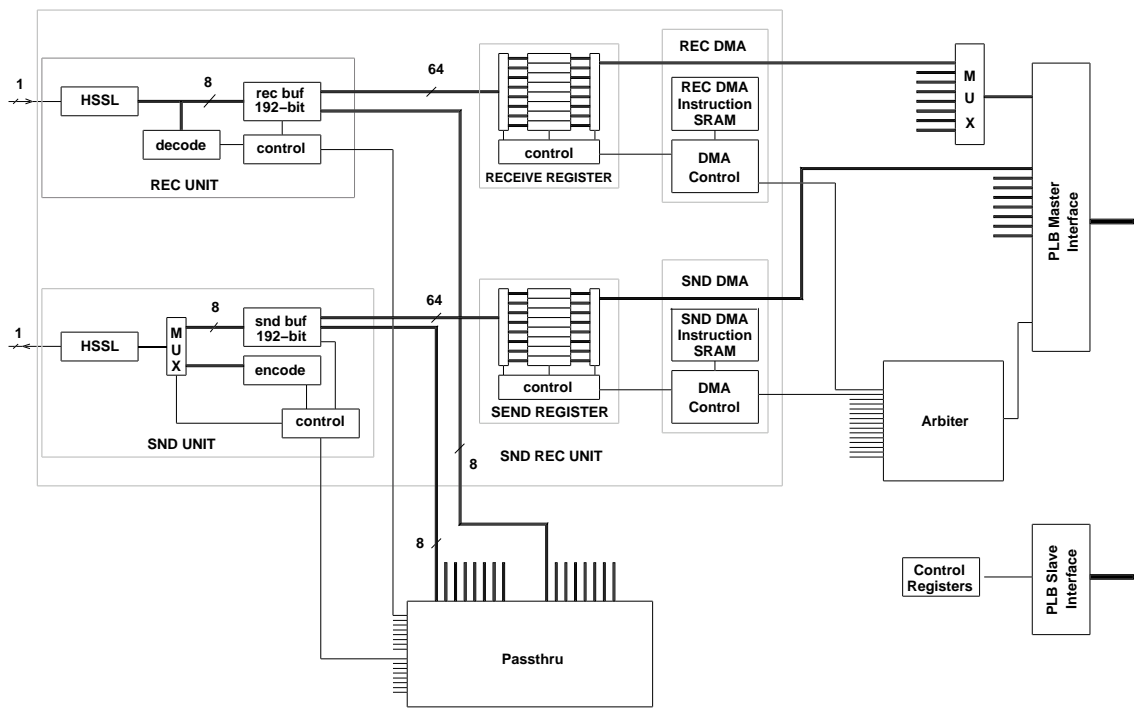}
    \caption{Serial Communications Unit in the QCDOC ASIC.  The ASIC
      boundary is on the left, the PLB interface on the right.}
    \label{fig:scu}
  \end{figure*}

\begin{itemize}\itemsep 0mm
\item Serial Communications Unit (SCU).  The task of the SCU is to
  reliably manage the exchange of data between neighboring nodes with
  minimum latency and maximum bandwidth.  The design takes into
  account the particular communication requirements of lattice QCD
  simulations.  A schematic picture of the SCU is shown in
  Fig.~\ref{fig:scu}.
  
  The custom protocol governing the data transfers defines packets
  that contain a 64-bit data word and an 8-bit header containing
  control and parity bits.  When the receive unit receives a packet,
  it first interprets the header, buffers the bytes from the HSSL, and
  assembles the 64-bit word.  It then transfers the word to the
  receive register or passes it on to the send unit.  The receive
  buffer can store three 64-bit words so that the send unit (in a
  neighboring node) can send three words before an acknowledgment is
  received.  The functionality of the send unit is essentially the
  inverse of that of the receive unit.  Send and receive operations
  can proceed simultaneously.  Each send or receive unit is controlled
  by a DMA engine which then transfers the data between memory and a
  send/receive register.  Each DMA engine is controlled by
  block-strided-move instructions stored in SRAM in the SCU itself.
  
  A low-latency passthrough mode is provided for global operations.
  Because of the latencies associated with the HSSL, the most
  efficient method to perform global sums is ``shift-and-add'', using
  a store-and-forward capability built into the SCU.  The main
  advantage of this scheme is that the software latency of about 300
  ns is paid only once per dimension, rather than for each node in
  this dimension.
  
  The total end-to-end latency is estimated to be about 350 ns
  for supervisor transfers and about 550 ns for normal transfers.
  This is at least an order of magnitude lower than the latency
  associated with Myrinet.  Since a write instruction from the 440 can
  initiate many independent transfers on any subset of the 24 send or
  receive channels, the latencies associated with multiple transfers
  can be overlapped to some degree.  The total off-chip bandwidth
  using all 24 HSSL ports is 12 Gbit/s.  In a 4-dimensional physics
  application only 16 of the 24 HSSL ports will be used, resulting in
  a total bandwidth of 8 Gbit/s.  This provides a good match for the
  communications requirements of typical applications.  Concrete
  performance figures are given in Sec.~\ref{PeterPerformance} below.
  
\item Prefetching EDRAM Controller (PEC).  The PEC is designed to
  provide the 440 with high-bandwidth access to the EDRAM.  It
  interfaces to the 440 data read and data write busses via a fast
  version of the PLB that runs at the CPU frequency and that we call
  processor direct bus (PDB).  The PEC also contains a PLB slave
  interface to allow for read and write operations from/to any master
  on the PLB as well as a DMA engine to transfer data between EDRAM
  and the external DDR memory.
  
  The access to EDRAM (which is memory-mapped) proceeds at 8 GBytes/s.
  ECC is built into the PEC, with 1-bit error correct and 2-bit error
  detect functionality.  The PEC also refreshes the EDRAM.  The
  latency of the PDB itself is 1-2 CPU cycles.  This very low latency
  eliminates the need for pipelining.  The maximum PDB bandwidth is 8
  GBytes/s for read and write.  However, due to internal latencies in
  the 440, the maximum sustained bandwidth is 3.2 GBytes/s.
  
  The read data prefetch from EDRAM occurs in two 1024-bit lines.
  Three read ports (PDB, PLB slave, DMA) arbitrate for the common
  EDRAM.  The coherency between PDB, PLB slave, and DMA is maintained
  within the PEC.  Each read port has four 1024-bit registers that are
  paired in two sets to allow for ping-ponging between different
  memory locations.  There are also two 1024-bit write buffer
  registers each for the PDB/PLB slave/DMA write interfaces.
  
\item Ethernet-JTAG interface.  As mentioned above, the 440 core has a
  JTAG interface over which one can take complete control of the
  processor.  In particular, this interface can be used to load boot
  code into the instruction cache and start execution.  This
  completely eliminates the need for boot ROM.  The question is how
  the JTAG instructions should be loaded into the 440.  (There are
  special tools that use the JTAG interface, but it would be
  impractical to connect one tool per ASIC for booting.)  A solution
  to this question has already been developed at IBM Research,
  implemented using a field-programmable gate array (FPGA), that
  converts special Ethernet packets to JTAG commands and vice versa.
  This logic is now part of the QCDOC ASIC and will be used not only
  for booting but also to access the CPU for diagnostics/debugging at
  run time.  The unique MAC address of each ASIC is provided to the
  Ethernet-JTAG component by location pins on the ASIC, and the IP
  address is then derived from the MAC address.
\end{itemize}

The QCDOC ASIC is manufactured using CMOS 7SF technology (0.18 $\mu$m
lithography process).

\subsection{Mechanical design}

\begin{figure}
  \includegraphics[angle=270,width=80mm]{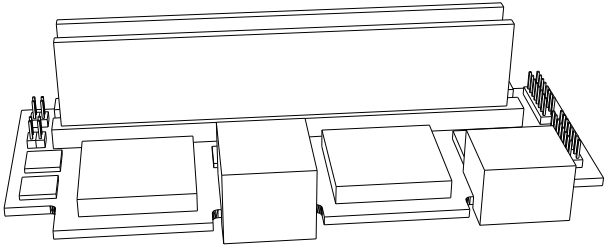}
  \caption{A QCDOC daughterboard with two ASICs and two DDR DIMMs.
    The cubic blocks are connectors.}
  \label{fig:dbd}
\end{figure}

\begin{figure}
  \includegraphics[angle=90,width=80mm]{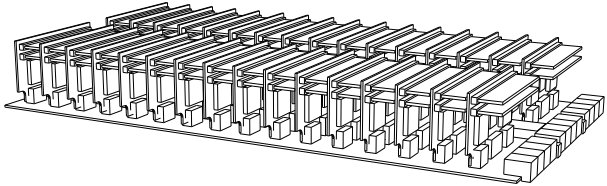}
  \caption{A QCDOC motherboard with 32 daughterboards.}
  \label{fig:mbd}
\end{figure}

\begin{figure}
  \includegraphics[angle=270,width=65mm]{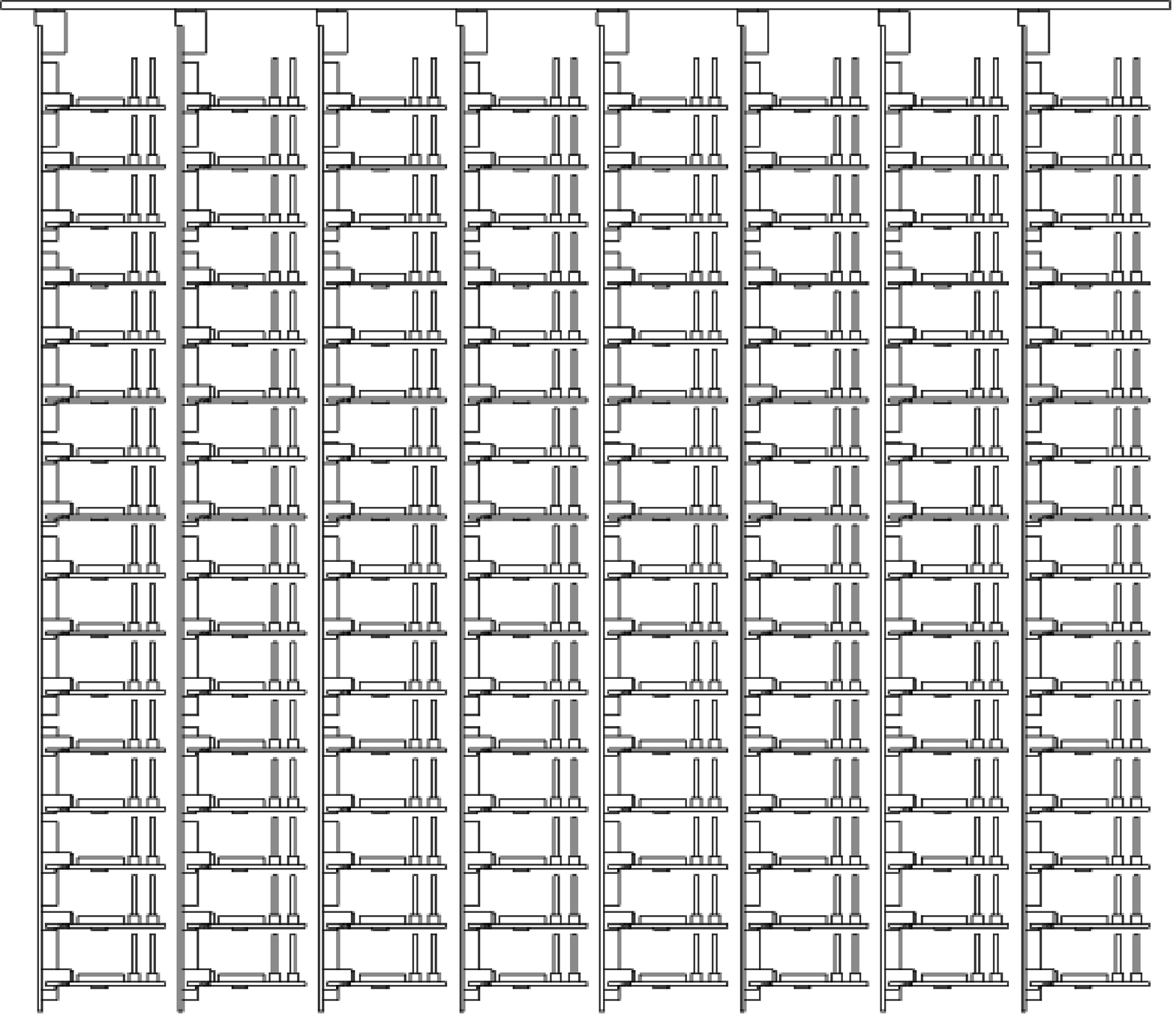}
  \caption{8 QCDOC motherboards in a single backplane.}
  \label{fig:backplane}
\end{figure}

The mechanical design is indicated in
Figs.~\ref{fig:dbd}--\ref{fig:backplane}.

Two QCDOC ASICs will be mounted on a daughterboard, together with two
industry-standard DDR SDRAM modules (one per ASIC) whose capacity will
be determined by the price of memory when the machine is assembled.  A
maximum of 2 GBytes per ASIC are supported.  The daughterboard also
contains four physical layer chips (two per ASIC for their Ethernet
and Ethernet-JTAG interfaces) and a 4:1 Ethernet hub so that a single
100 Mbit/s Ethernet signal is taken off the daughterboard.

32 daughterboards are mounted on a motherboard.  The motherboard also
contains eight 4:1 Ethernet hubs so that the total Ethernet bandwidth
off a motherboard is 800 Mbit/s.  Furthermore, the motherboard
contains power transformers as well as temperature and voltage
sensors.

8 motherboards are mounted in a crate with a single backplane.  The
final machine then consists of a certain number of such crates
connected by cables.

\subsection{Networks}

There are three separate networks: the high-speed physics network, an
Ethernet-based auxiliary network, and a global interrupt tree.  

The physics network consists of high-speed serial links between
nearest neighbors with a bandwidth of 2$\times$500 Mbits/s per link.
Transfers across these links are managed by the serial communications
unit in the QCDOC ASIC as described above.  The nodes are arranged in
a 6-dimensional torus which allows for an efficient partitioning of
the machine in software, described in more detail below.  On a
motherboard, the node topology is $2^6$, with three dimensions open
and three dimensions closed on the motherboard (one of which is closed
on the daughterboard).

The auxiliary network is used for booting, diagnostics, and I/O over
Ethernet, with an Ethernet controller integrated on the ASIC.  The
Ethernet traffic to and from the ASIC will run at 100 Mbit/s.  As
mentioned above, hubs on the daughter- and motherboards provide a
total bandwidth of 800 Mbit/s off a motherboard to commercial
Gbit-Ethernet switches, a parallel disk system, and the host
workstation (a standard Unix SMP with multiple Gbit-Ethernet cards).

The global interrupt tree consists of three separate interrupt lines
that are visible across all partitions.  An interrupt asserted by any
ASIC is first transmitted to the top of the tree and then propagated
down the tree to all other ASICs in the full machine.  It will remain
asserted until cleared by the ASIC from which it originated.

\section{System Software}

The QCDOC system software is minimally required to 
boot and manage the machine, load and run application code, 
and service application, communication and I/O requests.
The system can be thought of as composed of three major parts.

\begin{itemize}\itemsep 0mm
\item The front-end operating system, known as the qdaemon. 
\item The node operating system, known as the run-kernel.
\item The run-time support libraries used by applications.
\end{itemize}

In this section we discuss the qdaemon and run-kernel and defer
discussion of the run-time support libraries till a later
Sec.~\ref{ChulwooSciDacAndSCU}.

\subsection{qdaemon}
The qdaemon runs like a normal Unix daemon and is responsible for
booting, managing and partitioning the "back-end" grid of QCDOC nodes.
The qdaemon is the sole point of access to the back-end for users and
communicates with many nodes concurrently via RPC over the multiple
gigabit Ethernet links.

The qdaemon is contacted either via a PBS based queuing system, or
directly via a client program called the "qcsh". The qcsh is a
modified shell, with additional built-in commands for sending requests
to the qdaemon to perform operations on the QCDOC, such as running
code on a partition.

The qdaemon is aggressively multi-threaded and supports many
partitions and connected users simultaneously.

\subsection{Run-Kernel}

The node operating system is a simple (non-preemptive) kernel.  The
overall design goal for the run-kernel is to run one compute process,
and run it well.

Thus the kernel uses the 440's MMU for memory protection, but not
translation. This allows for protection of the O/S from errant user
code, and for zero-copy communication with simple (i.e. non-virtual
memory aware) hardware.  Further, the entire memory map can be covered
by the 64 entry TLB, so that TLB misses (which are a source of
significant performance loss in many HPC machines) cannot occur on
QCDOC.

The kernel does not implement scheduling so that the application is
guaranteed 100\% of the CPU.  This is very important since a more
traditional kernel on such a large and \emph{very} tightly coupled
machine (\mbox{QCDOC} will self-synchronise every 22 microseconds in
some codes) would be impacted by a few nodes running their scheduler
during any given dslash application.

The run kernel does, however, service both hardware and software
interrupts.  The features of the standard PowerPC architecture allow
user code errors to be trapped cleanly and robustly from software
interrupts.  The kernel also services system call requests from user
code, both to access the communication hardware through a very lean software
layer, and to implement the standard C run-time environment (Cygnus newlib).

The kernel includes an Ethernet driver allowing for host-node
communication, and an NFS client has been implemented allowing for
file I/O from the nodes, both to the front end and to a parallel disk
system composed from standard network attached storage.

Both run-kernel and application circular print buffers are maintained
on the nodes. This allows for post-mortem readout of each node's
output, or optionally any subset of nodes can be configured at
run-time to output directly to the console.

\subsection{Partitioning}

As discussed, QCDOC is based on a six-dimensional toroidal grid of
nodes.

A partition is a rectangular subvolume of the total machine and can be
defined by the 6-coordinates of two nodes in the machine grid, namely
the bottom-left, upper-right pair in six dimensions.

Each node has its own mapping from application directions to machine
directions. By changing this mapping with node coordinate, we can
change the topology and dimensionality seen by application code. This
is done by successively folding machine axes together into a single
application axis.

\begin{figure}[-b]
\includegraphics[width=6cm]{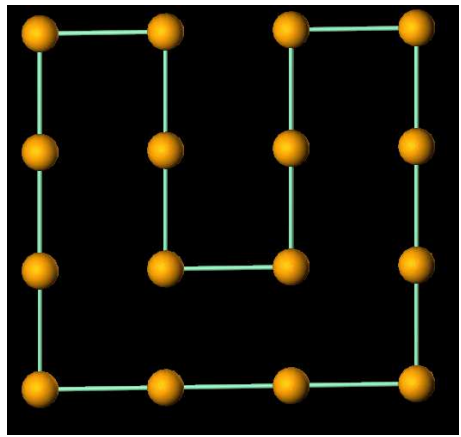}
\caption{ \label{Fig4x4}
Folding two machine axes into one periodic application axis}
\end{figure}

A simple 2d example of mapping an in principle non-periodic $4\times4$
square into a 1d-torus of length 16 is shown in Fig.~\ref{Fig4x4}.

\begin{figure}[-t]
\includegraphics[width=6cm]{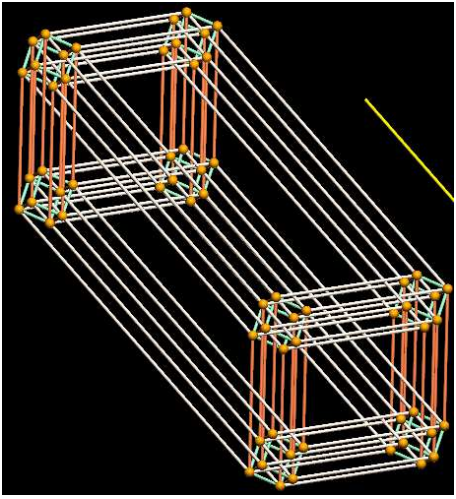}
\caption{ \label{Fig4x4x4}
Folding three pairs of machine axes into three periodic application axes.
This example corresponds to remapping a single QCDOC motherboard.}
\end{figure}

This process may be repeated, and an example of remapping a $2^6$
single motherboard is shown in Fig.~\ref{Fig4x4x4}. For any given
machine it is possible to configure \mbox{QCDOC} for any
application dimensionality from 1 to 6. From the point of view of the
communications software, all partitions are six dimensional, but some
variable number of these dimensions may be trivial. Calls in trivial 
application dimensions implement local copies from a node to itself,
while calls in non-trivial application dimensions are mapped to the appropriate
machine directions.

\subsection{Performance of optimised code}
\label{PeterPerformance}

In this section we present the performance of optimised code on cycle
accurate simulation of QCDOC.  Table~\ref{TabFPU} shows the
performance of key single-node assembler kernels.  Most of these
assembler kernels were in fact generated via a C++ program which will
also output Alpha and Sparc assembler. This allows a very fair
comparison with other contemporary RISC chips.

\begin{table}[-b]
\caption{
\label{TabFPU}
Double precision floating point performance on optimised variants
of common (single-node) QCD kernels on the QCDOC simulator. 
Here we assume a 500MHz nominal clock, and 
very high fractions of the 1 GFlop/s peak can be obtained in the dominant
kernels such as SU3-2spinor.
}
\begin{tabular}{c|c}
Operation & Mflops/node \\
\hline
SU3-SU3      &  800\\
SU3-2spinor        &  780\\
DAXPY              &  190\\
ZAXPY              &  450\\
DAXPY-Norm         &  350\\
CloverTerm/asm     &  790\\
CloverTerm/gcc     &  150\\
CloverTerm/xlc     &  300
\end{tabular}
\end{table}

\begin{table}[hbt]
\caption{
\label{TabMultinode}
Performance of common QCD Dirac matrix multiply operations on the
QCDOC simulator. 
Wilson $D_{eo}$ takes around 22$\mu$s on $V_{\rm local} = 2^4 $,
with  16 distinct communications in that time.
}
\begin{tabular}{c|c|c}
Operation  & Local Vol. &  Mflops/node\\
\hline
Wilson    $D_{eo}$ & $2^4$ & 470\\
Wilson    $D_{eo}$ & $4^4$ & 535\\
Clover    $D_{eo}$ & $2^4$ & 560\\
Clover    $D_{eo}$ & $4^4$ & 590\\
Staggered $D_{eo}$ & $2^4$ & 370\\
Staggered $D_{eo}$ & $2^2.4^2$ & 430\\
Asqtad    $D_{eo}$ & $4^4$ & 440 \\
\end{tabular}
\end{table}

\begin{figure}[h]
\includegraphics[width=6cm]{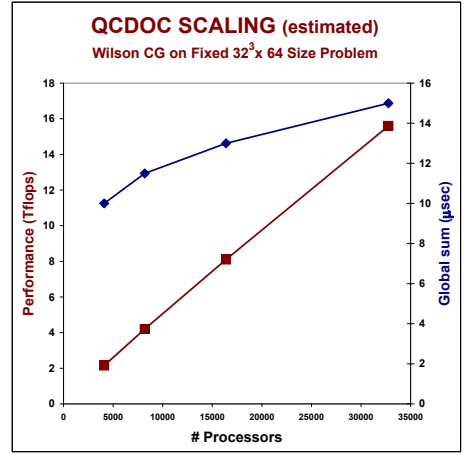}
\caption{ \label{FigPerf} 
Wilson CG performance on a fixed $32^3\times64$ lattice.
The machine should scale tremendously on a fixed problem size.}
\end{figure}

Anecdotally, the phenomenal bandwidth from EDRAM allows the FPU
performance of QCDOC on double precision floating point to match that
of high end workstations from L2 cache.

The very low latency communication on QCDOC (circa 550 ns) allows for
extreme scalability of QCD sparse matrix multiplies. Performance is
very acceptable down to local volumes as small as $2^4$, corresponding
to a very spread out lattice. These benchmarks are shown in Table
\ref{TabMultinode}.

Finally, we display estimates of the scalability of a CG Wilson solver
on large machines, including the 10 microsecond global summation time,
in Fig.~\ref{FigPerf}.

\section{User software and performance}
\label{ChulwooSciDacAndSCU}
User software can access the hardware capabilities of QCDOC by linking
to the run-time library provided.  Optimized lattice QCD computation
kernels as well as communication routines will be available.  The
software infrastructure, including run-time library and data
management and execution environment, will be in compliance with the
SciDAC lattice QCD software effort to ensure greater portability and
efficiency for a wide range of codes not necessarily optimized for the
QCDOC hardware.

\subsection{SciDAC software Program}
The SciDAC National Computational Infrastructure for Lattice Gauge
Theory initiative \cite{SciDAC} is a multi-institutional effort to
create the software and hardware infrastructure needed for Teraflop
scale lattice QCD simulations.

The goal of the software effort is to create a unified programming
environment that will enable the US lattice community to achieve high
efficiency on diverse multi-Teraflop scale hardware platforms.  More
specifically, the SciDAC software effort aims at providing a run-time
environment for existing lattice QCD codes such as the Columbia
Physics System (CPS) \cite{CPS}, MILC code \cite{MILC}, or the SZIN
software system \cite{SZIN} so that they can be run and achieve high
performance on various hardware platforms by adopting the programming
environment provided by the SciDAC Software effort.  A major part of
this activity is to provide C and C++ API routines for lattice QCD
(QCD-API) on which we will focus in this report.  The performance of
MILC routines with QMP on the QCDOC simulator is also presented.

The organization of the QCD-API is as follows.
\leftmarginii 4mm
\leftmarginiii 4mm
\begin{itemize} \itemsep 0mm

\item Level 1: 
\begin{itemize}
\item QCD Message Passing API (QMP) : Inter-node communication 
\item QCD Linear Algebra API (QLA): Single node linear Algebra
\end{itemize}

\item Level 2: QCD Data Parallel API (QDP): Lattice-wide operations
  with both communication and computation.\\[2mm]
  Example: parallel transport\\
  $\quad\ \chi'(x) = U_{\mu}(x) \chi(x+\hat\mu) $ for all $x$ and
  fixed $\mu$, where $\chi(x)$ is the fermion field at position $x$
  and $U_\mu(x)$ is the SU(3) gauge field connecting $x$ and
  $x+\hat{\mu}$.

\item Level 3: Highly optimized routines for lattice QCD
\begin{itemize}
\item Dirac matrix inverter 
\begin{itemize}
\item Wilson, Clover
\item Staggered 
\item Asqtad (Improved staggered)
\end{itemize}
\item Hybrid Monte Carlo (HMC) Asqtad force term 
\end{itemize}

\end{itemize}

So far, many Level 3 QCD-API routines as well as QMP have been
optimized and implemented for \mbox{QCDOC}.

\subsection{Brief description of QMP}
QMP aims to provide portable, low-latency, high-bandwidth
communication routines suitable for lattice QCD.  Here we describe
some of the features of QMP. (For a more detailed description as well
as for the C and C++ binding of the QMP routines, see
Ref.~\cite{QMP_doc}.)

\begin{itemize}
\item Point-to-point communication
\begin{itemize}
\item Nonblocking (computation and communication can be overlapped)
\item Simultaneous, multi-directional transfer capability
\item Chained block/strided transfer capability
\item Separate routines for initialization and commencement of
  transfers $\rightarrow$ opened communication channels can be reused
  to minimize overhead for repeated transfers
\end{itemize}
\item Global operation
\begin{itemize}
\item Global reduction \\
  Global Sum, Maximum, Minimum operations for integer, single and
  double precision numbers are available, as well as general binary
  reduction
\item Broadcast
\item Barrier
\end{itemize}
\end{itemize}

\begin{figure*}[hbt]
\includegraphics[width=135mm]{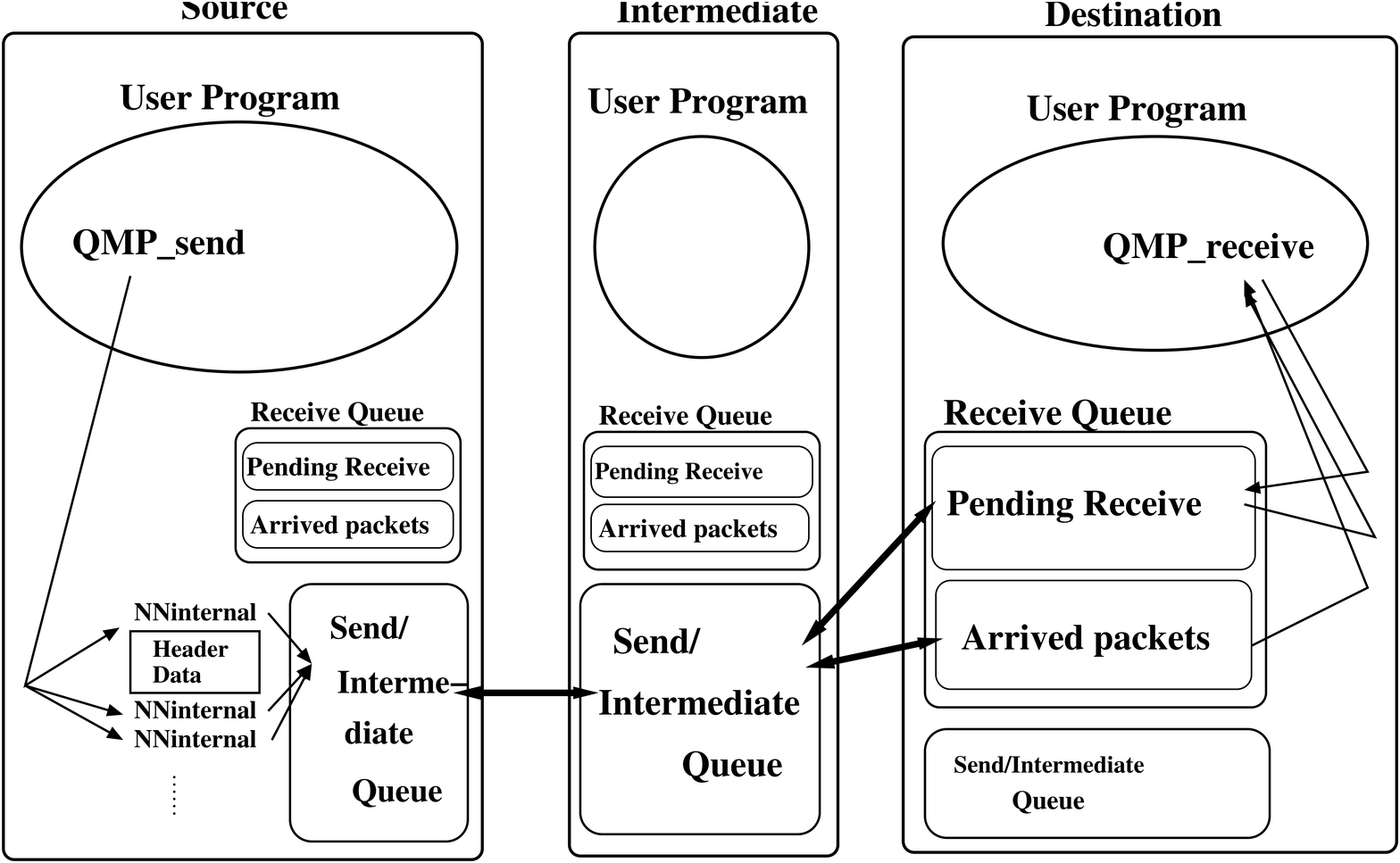}
\caption{\label{fig:NNC} Block Diagram of Non-Nearest Neighbor
  Communication QCD-API on QCDOC} 
\end{figure*}

Since QCDOC has optimized hardware for lattice QCD, it is crucial that
the implementation does not introduce an excessive software overhead.
For the nearest neighbor communication, the QMP specifications are
close to QCDOC native calls, and the additional software overhead is
small. Both C and C++ bindings are implemented over QCDOC system
calls.  This also applies to QMP global operations.  These are
implemented using the store-and-forward capability of QCDOC so that the
communication time for global operations grows only as $\sim L \times
N_d$, where $N_d$ is the number of dimensions and $L$ is the number of
sites per dimension.  Currently, the implementation of QMP is
complete except for non-nearest neighbor communication, which is
discussed in the next subsection.

\subsection{
  Non-Nearest Neighbor Communication (NNC) on QCDOC } While
communication patterns in lattice QCD are often to and from nearest
neighbors only, a well optimized NNC QCD-API can be used for an
implementation of routines with more complicated communication
patterns, such as improved discretizations of the continuum QCD
actions (Asqtad action \cite{asqtad}) and the MPI implementation on
QMP (which is under way \cite{MPI_QMP}). However, additional software
routines are needed to manage NNC on top of the QCDOC hardware where
all high-bandwidth connections are only to nearest neighbors.

Some of the desired features for NNC are as follows:

\begin{itemize}\itemsep 0mm
\item Minimal buffer copying.
\item Minimal overhead over QCDOC native communication call 
  when destination is nearest neighbor.
\item Capability to co-exist with nearest neighbor communication.
  Namely, both nearest and non-nearest communication can be open
  simultaneously and proceed without explicit user intervention or
  excessive delay on either communication channel.
\end{itemize}

The implementation strategy of NNC on QCDOC is as follows. After the
outgoing data is packetized in the source node, the header is
transferred and acknowledged via the interrupting communication
channel (supervisor).  The user data is then transferred using the
non-interrupting physics network.  If the receive node is not the
destination of the packet, the packet is pushed onto the
send/intermediate packets queue which sends the packets to the next
node on the path until it reaches the destination.  To avoid possible
lock-ups and excessive delays, static, Manhattan-style routing is
employed. It should be noted that in most lattice QCD routines, the
communication pattern is uniform relative to the source, in which case
Manhattan-style is close to being optimal. Figure \ref{fig:NNC} shows
the data flow between QCDOC nodes in NNC. The implementation of NNC is
under way \cite{MPI_QMP}.

\subsection{ Performance of MILC + QMP on QCDOC } 

MILC code \cite{MILC} is a body of C codes for SU(3) gauge theory,
developed and maintained by the MILC collaboration. It has been ported
to various parallel computers, workstations, and communication
protocols, including MPI and QMP.  Implementation of the QMP back-end
for the MILC code was done by James Osborne (osborn@physics.utah.edu).
We ran several lattice QCD kernels in MILC code with QMP on the QCDOC
simulator and measured the performance.

As shown in Table \ref{table:perf}, 15 $\sim$ 20\% of peak performance
is observed on the QCDOC simulator for many computationally intensive
MILC C routines in single precision compiled by the XLC compiler
\cite{XLC}, which generated significantly more efficient assembly
compared to GCC for PowerPC.  \hspace{-0.5cm}
\begin{table}[hbt]
\caption{\label{table:perf} Performance of QCD kernels from MILC code on
the QCDOC simulator. MILC codes are run in single precision, and a 500Mhz CPU
clock is assumed.}
\begin{tabular}{c|c|c}
Operation  & Local Vol. &  Mflops/node\\
\hline
Staggered $D_{eo}$ & $2^4$ & 170\\
Staggered $D_{eo}$ & $4^4$ & 210 \\
Asqtad    $D_{eo}$ & $4^4$ & 150 \\
Asqtad    Force    & $2^4$ & 140 \\
Asqtad    Force    & $4^4$ & 200 \\
\end{tabular}
\end{table}
\normalsize

\subsection{Improved action (Asqtad)  force term}

\begin{figure}
\includegraphics[width=80mm]{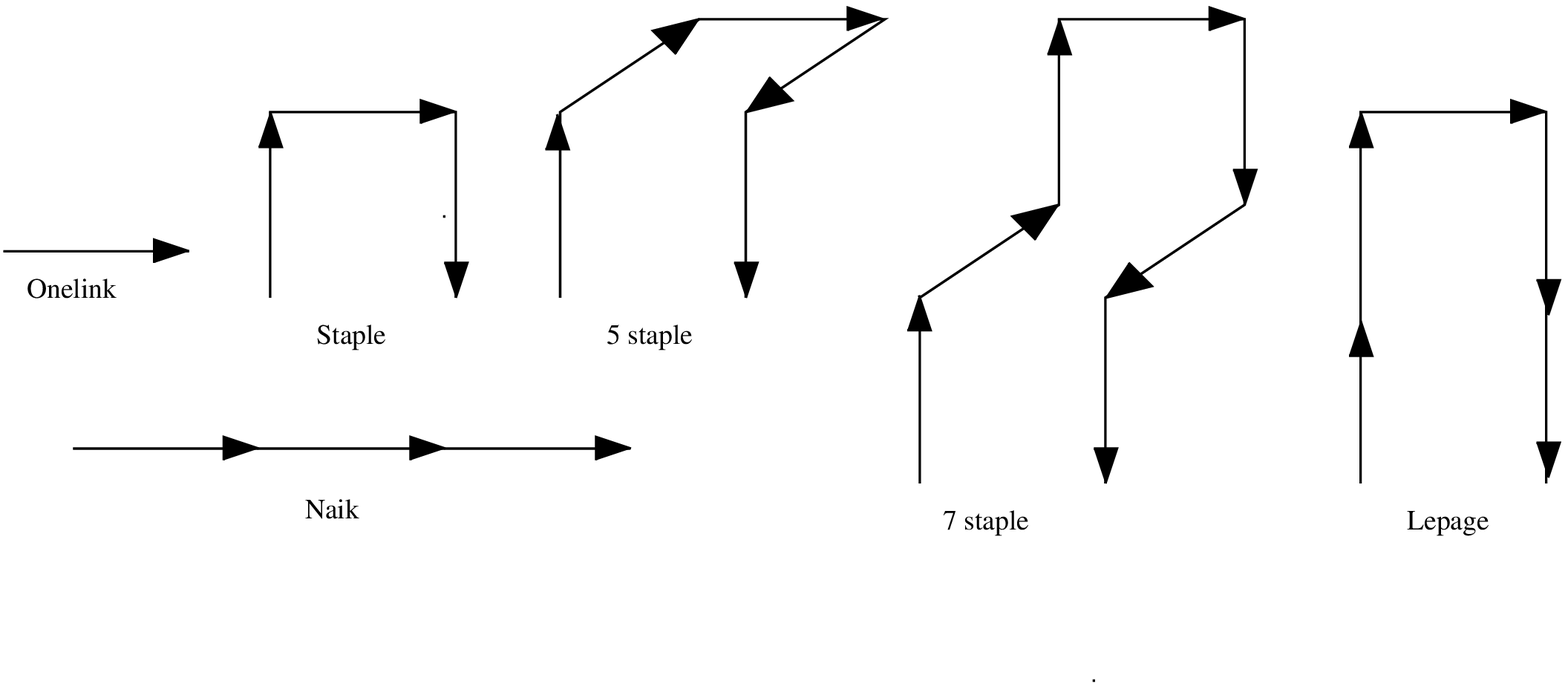}
\vspace*{-10mm}
\caption{\label{fig:asqtad}Diagram of paths used in the Asqtad action.}
\end{figure}

The Asqtad action \cite{asqtad} is one of the improved discretizations
of continuum QCD fermion actions which exhibits smaller lattice
spacing errors and flavor mixing. It uses many different paths
connecting $\chi(x)$ and $\chi(x+\hat{\mu})$ or $\chi(x)$ and
$\chi(x+3\hat{\mu})$ in contrast to the Wilson or staggered actions
where only one gauge link $U_{\mu}(x)$ is used for each pair of
neighbors.  A diagram of the paths in the Asqtad action is shown in
Fig.~\ref{fig:asqtad}.

Calculation of the force term for the gauge link $U_\mu(x)$,
${\partial S_{\rm Asqtad}}/{\partial U_\mu(x)}$, mostly consists of
parallel transport $ \chi'(x) = U_\mu(x) \chi(x+\hat\mu)$ and SU(3)
vector outer products $ P(x) = \chi(x){\chi'}^\dagger(x) $.  Because
of the complexity of the gauge paths used in the
action, the number of floating point operations for the Asqtad force
term is much larger than for the application of the Dirac operator and
the force term for simpler actions.  The number of floating point
operations per site is $\sim$ 250,000.  Considering that the number of
Flops for the Asqtad Dirac operator is only 1146 per site, and the
typical number of conjugate gradient iterations is about 1000 for the
mass used in typical lattice simulations, the force term calculation
is a significant portion of the Hybrid Monte Carlo simulation, where a
Dirac matrix inversion is alternated with a force term calculation,
and thus should be optimized as well as the Dirac operator.

We found the performance of the Asqtad force term was initially rather
low, only 3\% of peak for $2^4$ and 6\% for $4^4$ local lattice
volumes. Upon examining the source code, we found that the
communication channels are being created and destroyed for each
parallel transport.  This amounts to 30$\sim$60\% of the total number
of cycles for parallel transport.  After modifying the parallel
transport to reuse communication channels and combining small routines
used in the outer product routine to eliminate function call overhead,
the overall performance increased significantly to 12\% for a $2^4$
and 13\% for a $4^4$ local lattice.  

Further optimization was done by eliminating function calls for
computation routines defined for each lattice site within parallel
transport and outer product routines.  Together with loop unrolling,
this made it possible to preload data into the L1 cache and registers
to avoid cache misses.  This improved the performance of computation
routines by a factor of 1.5$\sim$1.7.  The overall performance
increased to 14\% for $2^4$ and 20\% for $4^4$ local volume, a
300$\sim$400\% increase over the original performance and quite
acceptable for a C routine.

\section{Status and conclusions}
QCDOC is a massively parallel computing architecture optimized for
lattice QCD.  The node has been designed to balance the floating point
performance, memory bandwidth, and communication performance such that
for QCD no single subsystem limits the performance.  The design
improves the network and memory subsystem performance relative to the
floating point peak when compared with current commercial MPPs.

Simulation measurements of the nearest neighbour latency and global
summation suggests at least an order of magnitude improvement over
current commercial machines. This has enabled us to demonstrate very
efficient use of the floating point unit that is maintained on
remarkably small local volumes, such as $2^4$, and estimate
scalability on typical lattices to machines as large as several tens
of thousands of nodes.

The six dimensional mesh network is dealt with transparently by the
operating system, and the machine can be dynamically partitioned to
run multiple applications of any dimension from one through six. This
should enable applications with similar characteristics to QCD (local
communication on a regular multi-dimensional mesh) to make efficient
use of the machine.

A run-time environment compliant with the SciDAC software effort will
be available.  The message passing interface defined by the SciDAC
effort, QMP, provides portable, efficient communication routines for
lattice QCD.  The QMP implementation on QCDOC is complete except for
non-nearest neighbor communication, which is in the process of being
implemented.  Performance numbers for lattice QCD routines from MILC
code with QMP were presented.  To take full advantage of the hardware
capabilities of \mbox{QCDOC} provided via QMP, the user programs
should be written in a way that minimizes repetitive overheads, as
shown by the performance numbers for the Asqtad force term.

The first batch of QCDOC ASICs has been manufactured and is being
assembled for initial testing.  While we thank the editors for their
patience, we regret having been unable to further delay the submission
of these proceedings until after the hardware went online.
Performance figures obtained from real hardware will be available in
the near future.

\begin{acknowledgments}
  This work was supported in part by U.S. Department of Energy
  contracts DE-AC02-98CH10886 (CJ) and DE-FG02-91ER40608 (TW), by the
  RIKEN-BNL Research Center (TW), and by PPARC (PAB).
\end{acknowledgments}

\end{document}